\documentclass[sigconf, preprint]{acmart}
\usepackage{graphicx} % in preamble
\usepackage{ulem}
\usepackage{lipsum}
\usepackage{tcolorbox}
\newcommand{\new}[1]{\textcolor{black}{#1}}
\newcommand{\old}[1]{\unskip}
\newtcolorbox{mybox}{colback=red!5!white,colframe=red!75!black}

\AtBeginDocument{%
  }

\copyrightyear{2026}
\acmYear{2026}
\setcopyright{cc}
\setcctype{by-nc-nd}
\acmConference[CHI '26]{Proceedings of the 2026 CHI Conference on Human Factors in Computing Systems}{April 13--17, 2026}{Barcelona, Spain}
\acmBooktitle{Proceedings of the 2026 CHI Conference on Human Factors in Computing Systems (CHI '26), April 13--17, 2026, Barcelona, Spain}
\acmPrice{}
\acmDOI{10.1145/3772318.3793706}
\acmISBN{979-8-4007-2278-3/2026/04}

\begin{document}

\title{Exploring Student Feedback Needs and Design Opportunities in Data Storytelling Education}


\author{Jennifer Posada}
\affiliation{%
  \institution{Department of Information Systems, University of Maryland, Baltimore County}
  \city{Baltimore}
  \state{Maryland}
  \country{United States}}

\author{Taha Hassan}
\affiliation{%
  \institution{Department of Computer Science, University of Alabama}
  \city{Tuscaloosa}
  \state{Alabama}
  \country{United States}}

\author{Lujie Karen Chen}
\affiliation{%
  \institution{Department of Information Systems, University of Maryland, Baltimore County}
  \city{Baltimore}
  \state{Maryland}
  \country{United States}}

\author{Louise Yarnall}
\affiliation{%
  \institution{SRI International}
  \city{Menlo Park}
  \country{California}}

\author{Jiaqi Gong}
\affiliation{%
  \institution{Department of Computer Science, University of Alabama}
  \city{Tuscaloosa}
  \state{Alabama}
  \country{United States}}

\renewcommand{\shortauthors}{Posada et al.}

\begin{abstract}
Data storytelling workflows ask learners to integrate analytical, design, and narrative skills, but instructors rarely have the capacity to provide detailed feedback at each step. Computational and AI-assisted storytelling offers opportunities to support student learning, but how feedback should be structured effectively remains unclear. To address this gap, we conducted a two-phase participatory design study. Through participant observations (N=8) and interviews (N=6), the first phase explored learners and educators' feedback needs and challenges in a data storytelling course. The second phase conducted two design workshops (N=8/10) to design and evaluate feedback strategies (frequency, seamlessness, accountability) for Story Studio: an AI-assisted narrative storytelling application. Our findings show that participants perceived on-demand and process feedback modes as effective, but automatic and outcome feedback as slightly more persuasive. We discuss implications for designing AI-augmented storytelling systems that adapt their feedback modes to the diverse needs and expectations of students.
\end{abstract}





\maketitle

\section{Introduction}
%Data storytelling presents key instructional challenges. Lack of relevant education research inhibits training learners on critical data storytelling skills that are in demand in the workforce. Limited time and resources to teach this skills makes scaling up this training difficult and there are opportunities to leverage AI to mitigate this, particularly in the area of instructor feedback. [Literature gap]

Data storytelling, also called narrative visualization \cite{segel2010narrative}, extends beyond traditional data visualization \cite{kosara2013storytelling}. It aims to effectively communicate data insights to stakeholders intending to prompt actions, a skill increasingly recognized as crucial for professionals with technical training, such as data scientists or analysts. Communication is explicitly listed as a core competency in various data science education frameworks, including reports from the National Academy of Sciences \cite{nasem2018undergraduates,nasem2020roundtable}
 and other data science and analytics competency frameworks \cite{leidig2020acm, demchenko2021edsf}. Training in data-driven communication skills or data storytelling is often a critical component of these frameworks. In practice, it may be integrated into the core curriculum of data science education under various names, such as data visualization, data-driven communication, or data storytelling. These topics may be offered as standalone courses \cite{nolan2021communicating} or as units in data science, analytics, or statistics courses. However, high enrollment in these courses presents challenges in providing timely personalized feedback to students as they develop their data storytelling skills, particularly when instructional resources are limited. In recent years, advances in AI—particularly Generative AI—have created opportunities to design a new generation of intelligent coaching systems capable of emulating expert human coaches. Such systems promise to deliver timely, targeted, and contextually appropriate feedback that can optimize learning. However, it remains unclear how these AI-driven supports can effectively meet the nuanced needs of students as they navigate the process of acquiring complex and ill-defined skill sets. 

%Using a participatory design approach, we engaged with student learners and experts through interviews, observations, workshops and a survey. We co-designed an intelligent coaching support platform and solicited feedback on iterations. Our findings indicate that users find AI assisted feedback to be effective for technical support and find it to be useful for learning given that there is a balance made between the level of agency and scaffolding provided for tailoring the narrative to their dataset and audience. This paper contributes 1) insights into the perceived efficacy of seamless feedback for AI-assisted data storytelling education and 2) and understanding of how much scaffolding users consider helpful for learning. 3) a design artifact for an intelligent coaching support platform OR implications for design. 

Using a participatory design approach, we engaged student and instructors/experts through interviews, observations, workshops, and an evaluation survey. We co-designed an intelligent coaching support platform and gathered feedback across multiple iterations. Our findings suggest that users perceive AI-assisted feedback as effective for technical support and valuable for learning—particularly when an appropriate balance is struck between learner agency and scaffolding, enabling them to tailor narratives to both their dataset and audience.

This paper aims to answer two research questions:

\begin{itemize}
\item \textbf{RQ1}: What are learners and educators' needs and expectations of feedback on data storytelling workflows?

\item \textbf{RQ2}: What is the efficacy of user-preferred feedback modes for AI-assisted data storytelling tools?
\end{itemize}

This paper contributes:
\begin{itemize}
    \item Empirical insights into the perceived efficacy of seamless AI-assisted feedback to support developing data storytelling skills.
    \item An understanding of how much scaffolding learners consider beneficial for supporting their data storytelling workflows.
    \item Design implications and a concrete artifact — an intelligent coaching support platform — to guide future development of AI-driven educational tools.

\end{itemize}

\section{Related Work}

\subsection{Data Stories and the Data Storytelling Process}

Storytelling is a powerful tool for human communication; however, its value in data science has yet to be fully realized. Data storytelling refers to the skills required to communicate effectively with quantitative information through data, visualization, and narrative, seamlessly woven together to deliver a coherent and compelling communication product. The related research has its roots in information visualization, data journalism, and business analytics practices, but its study within data science education has only begun to emerge in recent years. Data stories take many forms, but the most commonly seen belongs to the genre of ``slide show” \cite{segel2010narrative}
which entails a sequentially ordered common data visualizations, such as scatter plots and bar charts, organized to present a logical progression of a ``storyline” and accompanied by a narrative, working together to deliver a coherent and persuasive message. This study focuses on this genre given its popularity in data science and analytics education. 

Data storytelling involves a complex and iterative cognitive process that combines quantitative reasoning with the creative task of crafting compelling visual narratives. It begins with a set of data insights (e.g., data plots) derived from the data science workflow and proceeds through an iterative and interactive process that transforms those “raw material” into stories in various formats. Our work draws from literature detailing the data storytelling authoring process. For example, Lee and authors \citep{lee2015story} outline a three-step storytelling process involving “explore data” (gathering facts), “make a story” (finding a storyline, organizing story pieces, and establishing logical connections), and “tell a story” (building and sharing the presentation). In this research, we focus on the “make a story” phase, or story construction, which is often seen as a mysterious and creative endeavor \cite{showkat2021stories} thus presenting significant challenges for beginners. To understand the cognitive processes involved, we use the framework proposed Noland \& Stoudt in \cite{nolan2021communicating}, which breaks story construction into concrete steps, such as grouping, ordering, filtering, and adding narrative. 
Storyboarding is a planning tool commonly used in creative fields like film and animation. It provides a structured approach to document how a story may unfold before committing resources. Data storytelling professionals frequently adopt storyboarding techniques \cite{dykes2019storytelling} to plan narratives during the story construction process. A data storyboard is defined as a “visual outline for presentation content” \cite{knaflic2022storytelling} and serves as a tool to organize thoughts into a coherent and compelling story \cite{wang2025jupybara}. Storyboarding supports critical decisions such as filtering data, determining sequence, identifying gaps, and gaining a holistic view of the narrative structure \cite{nolan2021communicating}. \new{There are various tools in this space to support the data storytelling process \cite{Ren2023DataStorytellingNarrative,shi2020calliope,Zhao2021Chartstory,Wang2018Narvis} however, these focus more on providing the means to do the process or automatically producing stories and less on teaching how to do it.}

\subsection{Feedback in Educational Psychology and Learning Sciences}
Previous research in education, learning sciences and psychology has long examined feedback as a mechanism for supporting learning \cite{Wang2018PublicPeerReview}. 
Hattie and Timperley \cite{hattie2007feedback} describe feedback along the dimensions of “feed up”, “feed back”, and “feed forward”, emphasizing how feedback can guide learners toward goals, help them reflect on performance, and support planning of next steps. \new{Other work using their feedback model has looked at perceptions of feedback and the relationship to learning outcomes \cite{Cheah2020StructuredFeedback}. They show the importance of structured feedback as it relates to business communication student's report performance and how process-oriented feedback was seen as more useful than graded feedback. Although this was evaluated in context of a company supervisors providing feedback on student business reports, their findings can provide some support for the less explored area of providing feedback for data storytelling since business report writing shares important similarities. In addition, it underscores the importance of perception of feedback in the effectiveness of feedback similar to other work in the field of educational design \cite{BangertDrowns1991Feedback, Ilgen1979Feedback, Timmers2015MotivationFeedback}.
A robust body of literature demonstrates the effectiveness of formative feedback in improving the quality of students’ writing \cite{graham2011informing,macarthur2016instruction,panadero2023effects}. More recent work has examined the utility of AI-generated feedback in higher education. For example, Lo's work \cite{lo2025impact} reported findings from a randomized study showing that AI-generated feedback enhanced students’ writing quality.}
Shute \cite{shute2008focus} distinguishes between formative and summative feedback and highlights the role of “facilitative” versus “intrusive” feedback in scaffolding learning. Classical studies on feedback timing, such as Kulhavy and Anderson \cite{kulhavy1972feedback}, explore the effects of immediate versus delayed feedback on retention. Kluger and DeNisi’s Feedback Intervention Theory (FIT) \cite{kluger1996effect} emphasizes how feedback directs attention across task, self-regulatory, and motivational levels. These frameworks provide a strong foundation for understanding feedback, but they focus primarily on linear learning tasks, such as problem sets or writing exercises, and offer limited guidance for creative, multi-stage tasks in data storytelling. Feedback Intervention Theory \cite{kluger1996effect} suggests that feedback is most effective when it directs attention appropriately, with task-level feedback often producing better outcomes than self-level feedback. Work on self-regulated learning and learner autonomy underscores the importance of giving learners control to manage cognitive load and sustain engagement.  \new{For example, Wang et al. \cite{Wang2018PublicPeerReview} emphasize that feedback with too much detail may affect motivation to address feedback. } Similarly, research on flow and interruptions \cite{horvitz1999disruption} shows that the timing and delivery of feedback can either disrupt or support engagement depending on whether it aligns with the learner’s workflow. These psychological models have rarely been evaluated for creative, iterative, and multimodal tasks like data storytelling, leaving open questions about how different feedback designs affect reflection, engagement, and learning outcomes.

\subsection{Feedback in Creative Storytelling}
HCI investigations into creativity and storytelling highlight reflection, multi-step non-linear task flows, and iterative decision-making as central to complex tasks. Schön \cite{schon1983reflective} distinguishes between reflection-in-action and reflection-on-action, showing the importance of both midstream and retrospective reflection for supporting creative problem solving. Csikszentmihalyi \cite{csikszentmihalyi1996creativity} emphasizes that creative work often alternates between exploratory and convergent phases. 
%Data storytelling is the construction of stories with visual, data-driven evidence and coherent narrative structures. A unique project within creative storytelling, it marries the evidence-gathering processes of quantitative research with narrative construction and critique. [Louise's design patterns]
Prior studies on creative cognition and iterative design workflows provide general guidance or scaffolds, but rarely structure feedback systematically according to task stage or type of reflection. This limits our understanding of how process-level versus outcome-level feedback can support learners as they make analytic and narrative decisions across multiple stages of a creative workflow. 

Research on seamfulness examines how system design exposes or hides complexity for users \cite{chalmers2004seamful}. Studies of intelligent tutoring systems and AI-assisted writing tools have largely provided task-specific, linear feedback in a single modality, without exploring how feedback can be tuned as an interaction mechanism across stages of a creative, multimodal workflow. Similarly, studies on interruptibility and attention management \cite{horvitz1999disruption} highlights how the timing of system interventions can influence engagement and reflection. Few studies, however, have applied these HCI perspectives to feedback design in complex creative tasks, leaving open questions about how interface design choices interact with learner agency, pacing, and reflection.

These studies collectively provide rich insights into feedback, attention, reflection, and creativity, but they primarily focus on linear or unimodal tasks. The multi-stage, creative learning activities involved in data storytelling remain to be addressed. Our study operationalizes three key feedback dimensions—requirement, timing, and locus—to empirically examine how feedback can be designed as an interactive, stage-sensitive mechanism that supports reflection, engagement, and creative decision-making. By bridging education, psychology, creativity, and HCI, our Story Studio project addresses a gap in guidance for designing feedback in complex creative workflows.

\section{Story Studio Project Overview}
\begin{figure*}[t]
  \centering
  \includegraphics[width=\textwidth]{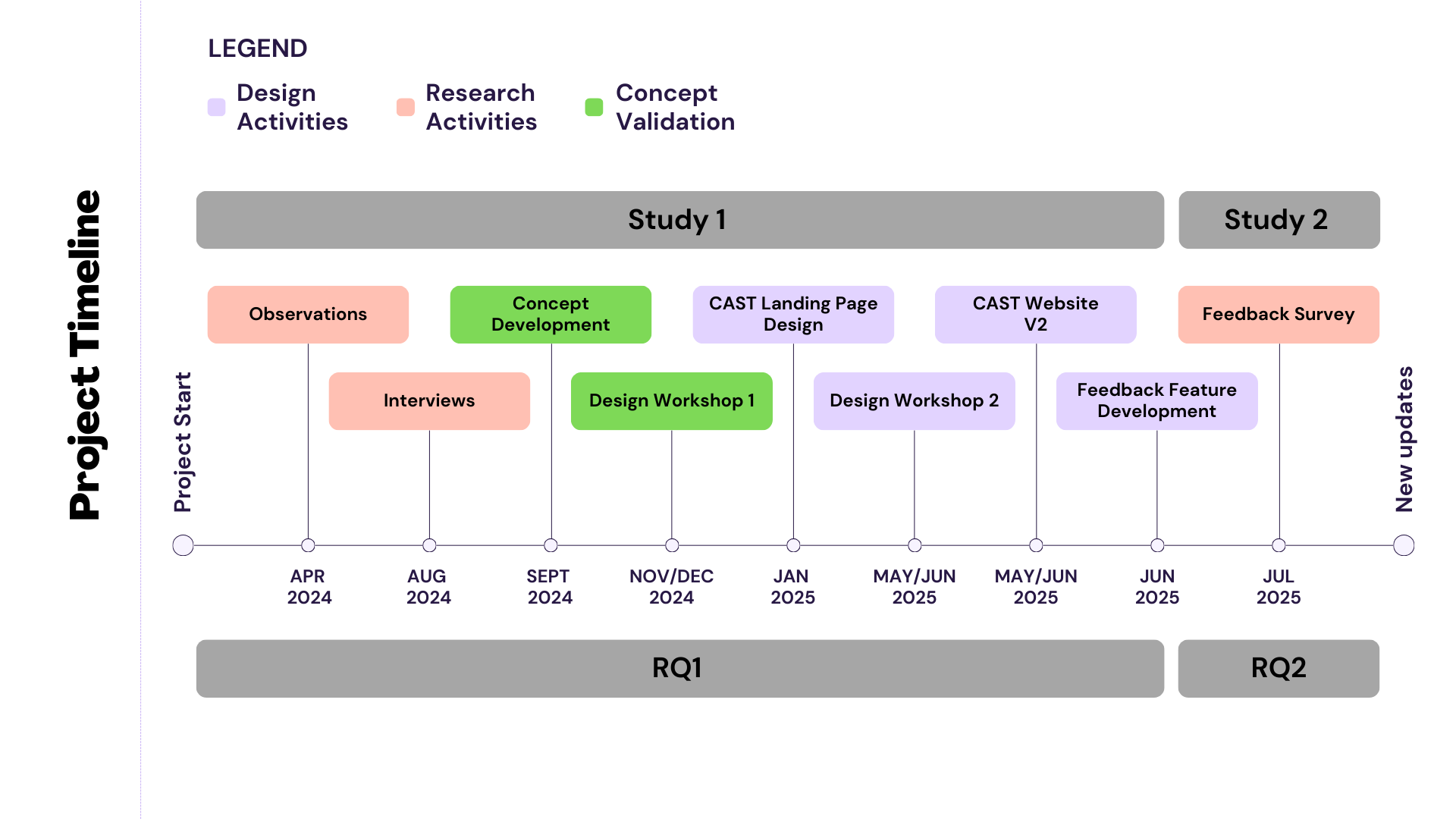}
  \caption{ Study timeline overview. (Top left). Study 1 activities (observations, interviews, concept development, and design workshops). (Top-right) Study 2 activities (feedback feature survey). Shows timeline of these activities and the corresponding study and research question number (bottom-left and bottom-right). }
  \Description{A visual showing study timeline overview. This timeline displays the different research activities conducted throughout the project lifecycle. It includes a legend distinguishing between design and research activities and concept validation. Activities are listed in chronological order from April 2024 to July 2025. It also displays overarching labels for which study and research question the activities corresponded to. }
  \label{fig:study_overview}
\end{figure*}

\begin{figure*}[h]
  \centering
  \includegraphics[width=\textwidth]{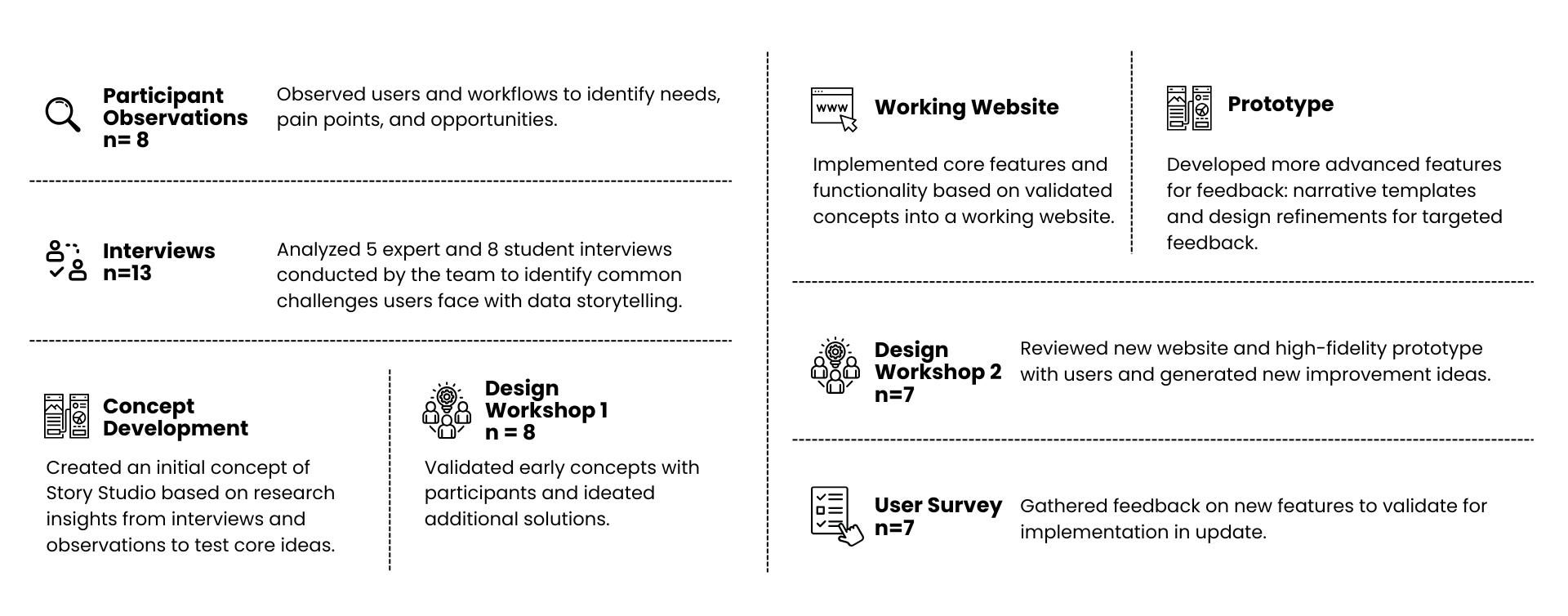}
  \caption{Study methodology overview. (Top left) Participant observations, interviews. (Bottom-left) Concept development, design workshop 1. (Top-right) Working website, prototype. (Bottom-left) Design workshop, user survey.}
  \Description{ Overview of the study methodology. . In the graphic there are 8 sections each titled, respectively, Participant Observation, Interviews, Concept Development, Design Workshop 1, Working Website, Prototype, Design Workshop 2, User Survey. Each shows more details about what was done in each phase of the study as discussed in Section 4.}
  \label{design_methods}
\end{figure*}

Figure \ref{fig:study_overview} presents the timeline of a series of participant design studies conducted to support the development of Story Studio, a scalable AI-empowered coaching platform that supports the development of data storytelling skills as a complement to existing data science and analytics education at the post-secondary level. 
Figure \ref{design_methods} provides a methodology overview of the specific design activities undertaken in each study.
\new{We employed purposive sampling to ensure a match between the recruited participants and the study aims and objectives of understanding learners and educators in the context of data storytelling.~\cite{Campbell2020PurposiveSampling}. }

The project comprises two phases. 
\begin{itemize}
    \item \textbf{Study 1} (Section \ref{section: study1}) is a multi-phase participatory design study that began with field observations of two small groups of undergraduate students in an introductory data science course engaged in storyboarding activities described by \cite{nolan2021communicating} in Spring 2024. This was followed by interviews with data science students, instructors, and data storytelling experts in Summer 2024 to understand their perspective in teaching and learning data storytelling. Based on these studies, we formulated design concepts and developed an initial prototype, which served as the starting point for two subsequent design iterations carried out through workshops and prototype development in Fall 2024 and Spring 2025. In Summer 2025, we refined the design of Story Studio, focusing on feedback features and synthesizing the insights accumulated up to that point. Due to space constraints, this paper focuses on the design workshop and the iterations, while providing a summary of the key findings from the observation and interview studies

    \item \textbf{Study 2} (Section \ref{section: study2}) involved a feedback survey with students and instructors, who evaluated the most recent iteration of Story Studio and provided input for further development.
    
\end{itemize}

\section{Study 1: Data Storytelling Needs and Expectations}
\label{section: study1}

\subsection{Study Objectives and Research Questions}
The main objective of this phase of the study is to identify the needs for learning and teaching support tools in data storytelling and, based on these needs, to develop a prototype that reflects them. Specifically, we aim to address the research question: \textit{What are learners and educators' needs and expectations of feedback on data storytelling workflows?} (RQ1)

\subsection{Study Methodology}
\subsubsection{Field Observation}

The main focus of this study is to understand learners’ struggles as they engage in story construction tasks—a challenging process that requires students to filter, group, and organize a set of data visualizations or data slices in order to craft coherent data stories.

\begin{figure}[t]
  \centering
  \includegraphics[width=\linewidth]{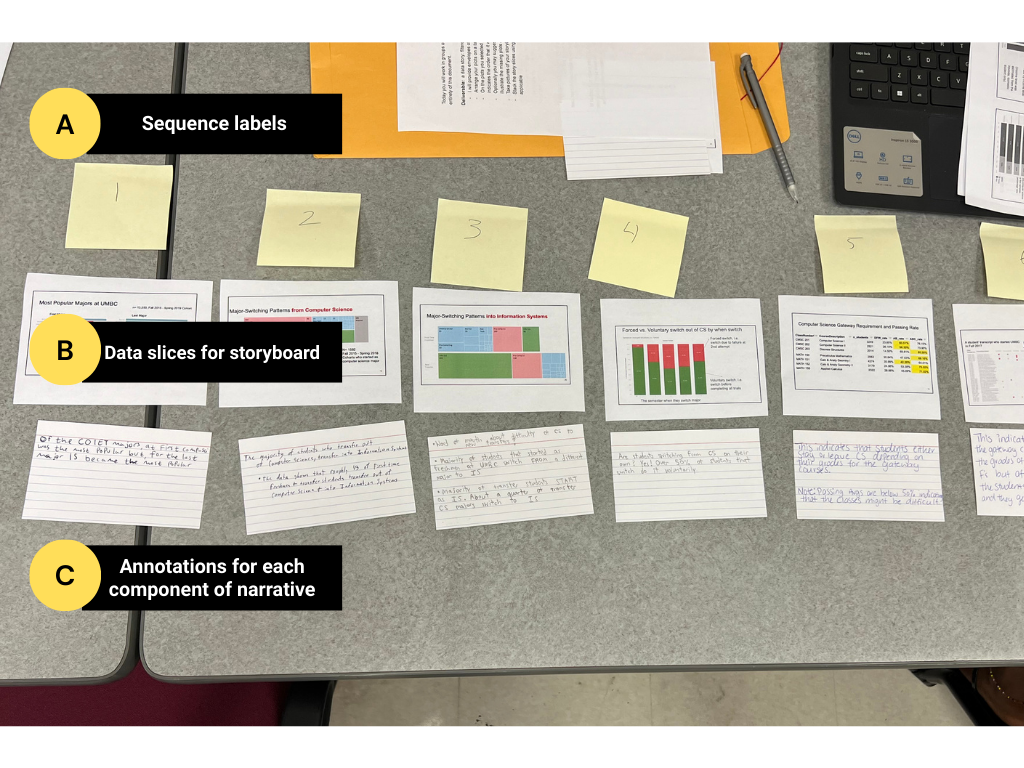}
  \caption{ Storyboard activity overview. (Top) This shows sticky notes at the top of workspace labeling the sequence of plots for the storyboarding activity, (Center) paper cutouts for data plots provided to students, (Bottom) and annotated cards the used as labels for their narrative}
  \label{Overview storyboard activity}
  \Description{This shows an overhead view of the students working on the storyboarding activity on a table in the classroom. It shows sticky notes that show the numbered order of the plot sequence. It also shows cutouts of data plots that students were provided with index cards underneath each plot with student written descriptions. }
\end{figure}

To investigate this process, we observed n = 8 students working collaboratively in groups of four on a storyboarding activity (Fig. \ref{Overview storyboard activity}) in an undergraduate introductory data science course in Spring 2024. \new{The observations provided a more detailed understanding of the moment to moment decisions students made during the storyboarding process in an education setting.} The students used data plots derived from a dataset on a topic relevant to their experiences. During these sessions, we collected audio recordings and took observation notes as students verbalized their thought processes, \new{and also collected student reflections about the activity.}
 which were later transcribed and coded for distinct sub-processes of data storytelling. 

\subsubsection{Semi-Structured Interviews with Instructors and Students}
We interviewed n = 5 experts with experience teaching data storytelling courses to better understand the key challenges they observed students facing. We asked these experts to identify the core knowledge, skills, and abilities (KSAs) associated with data storytelling. Their input served as a framework for understanding what learners need to accomplish in specific tasks and for guiding how the tool could be designed to support them. In addition, we interviewed n = 8 learners to explore the types of feedback they had received on data stories they previously created, as well as the kinds of feedback they would have liked to receive from instructors. We further inquired whether students had opportunities to act on the feedback and revise their stories, and what revisions they would have made to improve them. 

\subsubsection{Data Analysis and Early Concept Development}

Both the interviews and observations were analyzed and coded \new{using deductive coding with a predefined codebook} according to the three foundational design patterns for data storytelling \cite{yarnall2025laying}: (1) understanding the general audience and communication goals (understanding context), (2) data exploration and analysis (sensemaking), and (3) data story construction  \new{Since our research questions were related to these design patterns and the needs and expectations of feedback during the storytelling workflow, they served as predefined codes during our thematic analysis \cite{BraunClarke2006}. We started developing the codebook after reviewing one of the observations collaboratively and discussing initial results and any emerging sub-themes}. 
\new{For example, for learners' needs in storytelling workflow we found sub-themes around general problem-solving strategies like overt help-seeking behaviors, initialization, what is next, and moment of stuckness, and reflection. The same analysis was used on the interviews and reflections students had written about the storyboarding activity. }

\new{The time each group spent in each phase was visualized on a Gantt chart (see \ref{Gantt Chart}) that was coded over time according to which state of the data storytelling process they were determined to be in \cite{yarnall2025laying,nolan2021communicating}. }

\new{The} time-series visualization of our observations revealed that students engaged in an iterative process focused on sensemaking and story construction. From the interviews, we found that experts most frequently observed students struggling with understanding context and constructing a coherent storyline. Learners also reported wanting support that was adaptive to their stage in the process—particularly in understanding context—and the ability to ask specific questions about the visualizations they were working with. \new{In addition, their reflections indicated that many had challenges around narrative construction and sequencing parts of their story.}.
\begin{figure*}[h]
  \centering
  \includegraphics[width=0.9\textwidth]{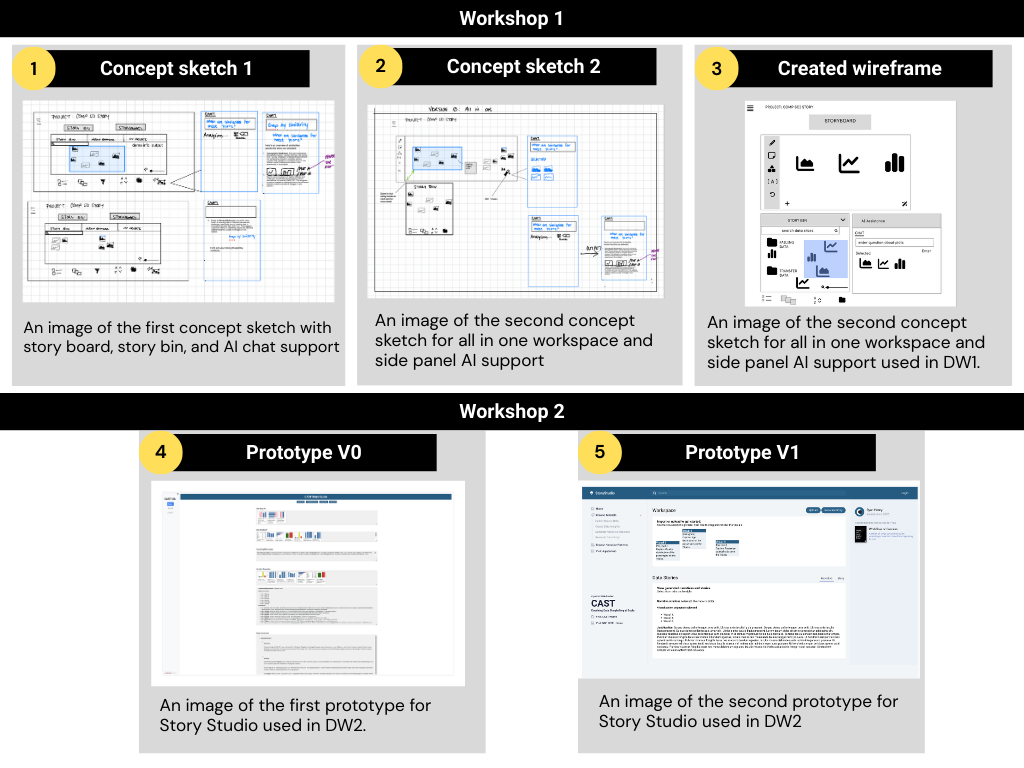}
  \caption{Conceptual design overview clockwise from left. (Top-left) Section 1 shows Concept sketch 1 which is the first concept with multi-stage workspace. Section 2 shows Concept 2 depicts an all in one workspace iterated from concept sketch 1 . (Top-right) Section 3 shows the wireframe created from conceptual design and used in the first design workshop. (Bottom-left) Section 4 shows Prototype V0 informed by co-design . (Bottom-right) Section 5 shows Prototype V1 used in DW2.  (Bottom-left and bottom-right) Concept wireframe 1 based on concept sketch 2.} 
  \Description{This image shows an overview of the process of creating the conceptual design. It shows 5 sections each showing a phase in the process, respectively, including: Concept sketch 1, Concept Sketch 2, Created Wireframe, Prototype V0 and Prototype V1. Each section shows a picture of the system at that phase.}
  \label{concept_sketch}
\end{figure*}
Based on these findings, we developed a conceptual design (Fig. \ref{concept_sketch}) for a platform that supports iterative interactions with digital data plots for data storytelling. The design incorporates both a storyboard and a story bin, drawing on metaphors consistent with those used by experts to describe their own processes. \new{We use an evidence centered design approach \cite{yarnall2025laying,Haertel2012ECD} in our concept development ensuring we tied the concept back to what students should learn or demonstrate during the storytelling process and barriers we wanted to mitigated}

The initial prototype (V0) was tested with eight participants. This prototype comprised several components: (1) a workspace for ordering data slices, which provided students with feedback on potential orderings to encourage exploration of different narrative perspectives; (2) features that enabled students to select and refine a storyline based on their intended audience and communication goals; (3) functionality for comparing alternative narrative structures, allowing students to evaluate which structure best suited their goals while also illustrating the range of perspectives the AI could generate; and (4) contextual feedback from the AI on student-selected storylines, offering targeted suggestions for refining their work. Findings from the first workshop motivated the development of prototype V1, which was tested in a second round of design workshops with seven participants. Insights from these workshops informed the creation of V2, which incorporated new feedback features. This version was subsequently evaluated in Study 2 with another seven participants

\subsubsection{Design Workshop I}

We conducted three co-design workshops with students enrolled in a data science course. Groups 1 and 2 each included three participants, while Group 3 included two participants. In the first workshop (Group 1), participants reviewed scenarios depicting two data analysts \new{named Jamie and Alex}—one working in the context of a streaming service and the other in healthcare—tasked with creating a data story. After selecting a scenario, participants were asked a series of guided questions, as detailed in Appendix \ref{design_workshop1}.
Working in a FigJam whiteboard space, participants brainstormed potential system features to support data analysts  and then presented their envisioned interactions. Afterward, we introduced the initial design iteration to gather feedback, identify sources of friction, and explore possible refinements. 

Based on observations from the first workshop, we modified the procedure for Groups 2 and 3. Several participants had shown unfamiliarity with the affordances of our concept and tended to default to a generative AI mental model, despite our initial decision to withhold generative data storytelling tools to avoid over-influencing their thinking. We speculated that this reliance was due to the novelty of the technology. To address this issue, we incorporated examples of existing data story creation tools \cite{wu2023socrates, shi2020calliope} for later groups, explicitly noting that our aim was to move beyond generative functionality toward more supportive features. The intention was twofold: (1) to encourage participants to conceptualize the system more concretely as a web-based tool, and (2) to familiarize them with the broader affordances of data story creation tools. These adjustments enabled Groups 2 and 3 to think more specifically about how they would expect the interface to function.

Participants generated solutions informed by both the demonstrated tools and the V0 prototype. They were asked to reflect on the challenges they had encountered in data storytelling and to identify the kinds of support they wished had been available. Insights from their interviews, in which they articulated desired system supports, further informed the process. Collectively, these findings guided the development of the next iteration (V1). 

\subsubsection{Design Workshop II}
The second round of design workshops were conducted virtually using Zoom and Figjam to understand user perceptions around the usefulness of V1 and how they envisioned feedback features to be implemented into the design. We showed them both the working website and a prototype draft for new visuals and functionality updates, asking what stood out and whether any parts were confusing or overwhelming. To understand their needs and expectations with feedback in their data storytelling workflow, we asked questions around format, timing, actionability and feedback source (see appendix Section \ref{design_workshop2} for questions). Each workshop for this round included one participant and one researcher. Solutions were ideated collaboratively with each student, where the researcher added their suggestions to stickies, and some students even used materials to prototype simple ideas. 

\subsection{Study Findings}

\subsubsection{Users Desire Situational Awareness, Scaffolding, Q\&A and Feedback in Storytelling Workflows}
The first round of design workshops surfaced several key themes around how participants envision their data storytelling tool experience. After reviewing the concept design we had, the first group expressed how they wanted some structure generated by the tool, an outline of which parts of the data were relevant to which audience, \new{and assistance for ensuring clarity is achieved with minimal jargon}. In addition, they noted they wanted a side panel to ask questions or a search bar to be consistent with current AI tools. The second group discussed wanting explanations for why certain visualization and narrative choices are recommended so they can learn from it. In addition, they agreed on counterfactual feedback and explanations for why not to use a specific narrative or visualization. They also expressed wanting AI assistance present throughout the process either embedded and accessed through a right click interaction or on the side of their workspace.

Additionally, participants agreed that they wanted the tool to ask them what part of the data storytelling process they were in so it could tailor feedback accordingly. They cited natural language interfaces as the most intuitive noting they thought being able to go back and forth with the assistant would be useful. In addition to natural language interaction they also wanted preset options to support varying levels of expertise where preset options could be for beginners and more advanced users can interact conversationally to discuss more specific refinements to their narrative. Finally, the last group presented features like asking the system about specific parts of the storyline and also the entire storyline. They wanted a feature that walked them through how to adjust the story to their audience and audience education level. Additionally, they stated wanting the system to ask them about their audience motivations so that the tool can recommend tailored narratives. In terms of their workspace, one participant mentioned they would like different options such as a whiteboard space or a staged view where each stage of the workflow shows you different tools.

These findings were synthesized and grouped according to participants needs for clarity, usability, and responsiveness. Participants wanted clarity around why certain choices are recommended for specific audiences to support their understanding of context. They expressed that they felt the system would be usable if it had natural language interaction with output in the form of step by step instructions and also preset options for AI feedback. Finally, they expressed needing responsiveness from the system in the form of 1) asking system questions throughout the process 2) the system adapting to their current workflow stage. We reviewed the solutions and needs derived during the workshop and used it to inform the changes for next iterations. First, for the natural language interaction participants requested, we presented output from AI that translates the narrative into recommended step by step ordering of data slices. Second, we integrated the first version of adaptive feedback, providing responsiveness in the form of stage awareness by incorporating a feature where the data slices placed in the storyboard section served as input for the LLM’s narrative feedback. 

\subsubsection{Users View AI Assistance in Identifying Narrative Structures Favorably}
In the prototype review, participants viewed the current website design along with an in-progress prototype and identified what features stood out to them. We identified users' needs for support in understanding the narratives to choose for their communication goals, the kind of AI feedback they envisioned receiving, and how they would interact with it.

During this session, we discussed areas for improvement based on their experience having the participants think aloud as they interacted with the website. Participants highlighted moments where they perceived AI feedback features to be generally useful including the AI generated narratives and the data plot descriptions as especially helpful. When reviewing the in progress prototype, most participants expressed being pleased with the new design highlighting the visual updates as more engaging. In addition, they felt that the narrative patterns features were useful and anticipated they would be helpful during their process. \new{One noted that there are a lot of elements to the narrative patterns and that AI could be particularly helpful in learning and understanding them. Another noted that they would find it especially helpful if AI provided a step-by-step visual of how a narrative pattern could be structured. This template style approach was deemed useful by most but some did note that they only wanted AI support to certain extent because they were "not going to read the extra details" suggesting the need to consider how much scaffolding is provided in this area.}

\subsubsection{Users Favor Stagewise, Instructor-Led Feedback, But Are Split on Actionability}
During the ideation session, the participants expanded on the observations made during the prototype and website review and refined potential solutions to improve the design. These solutions fell into two categories, feedback experience and narrative templates.

\textbf{Format}: They expressed that they found comments integrated within the workflow as annotations as being the most helpful format of feedback because it can provide context for what the feedback is about. One noted, ”\textit{comments are best…it depends if you want scoring it feels like a grade and get low percentage you might feel youve done bad work…but a sentence or two of how you could improve rather than just percentage}.” Among features like checklists, scores, and highlights participants felt that checklists were also helpful as it provided actionable steps to check off when addressing feedback. The use of visual cues like colored icons received mixed opinions with some participants highlighting its usefulness while others stated it could be distracting.

\textbf{Timing}: When discussing the best time to receive feedback, participants generally agreed that feedback in stages would be helpful to avoid overwhelm during learning. This overwhelm could occur if receiving negative feedback after investing substantial time in creating the story only to realize it was done incorrectly. One participant said, \textit{“I would prefer like a step by step process, like step by step feedback. And so it's not everything in a single place and overwhelming. It's just okay, you did it, you learn, you move. You did it, you learn, and you move.}”

They highlighted the need to balance feedback frequency with distractibility as too many feedback notifications might be unhelpful if it is distracting to some participants. This suggests the need to ensure scaffolding is not disruptive. 
\new{Discussions around timing also highlighted how it would be helpful to have opportunities to provide information about the audience in the beginning of the process. 
Others however, highlighted that they would prefer to keep feedback a one time thing particularly when it came to feedback on narrative structures.}

\textbf{Actionability}:  Participants noted preferences for how to take action on feedback given by the system, stating that they want clear steps to take action on but also open-ended feedback around how the story generally comes across. One gave an example of how they would like to receive an open ended feedback along the lines of “‘\textit{this is technically correct but you can do alot better.’..actionable steps should be available.}” They also want to be able to solicit clarification around the feedback given if it is still unclear expressing that an on-demand “ask question” feature for feedback would be helpful. \new{One participants said it "\textit{wouldn't hurt to have}" and compared it to other feedback they have received and how, "\textit{when they give you feedback you can't really ask [a question] because the feedback is pretty much you should understand it since it's based on what you wrote}". This suggests that the ability to ask for clarification around could mitigate the gap between the expectation of understanding feedback and the actual level of understanding.} When discussing features like revise now and mark as done, some participants cited how they would like to have agency over addressing the feedback with the option to skip the feedback\new{ particularly if they felt that have already addressed the issue}. Others expressed that they thought adding constraints for being able to proceed in the workflow would be helpful for learning to make sure they are going in the right direction. \new{One suggested that combining with automatic feedback would ensure "\textit{that they actually have to pay attention}" noting that when they know when to expect feedback "\textit{I kind of ignore it sometimes because I just want to get through whatever I am working on.}". This highlights the need to ensure actionability preserves agency while making sure that critical feedback is addressed before progressing. }

\textbf{Source}: Discussing which feedback they would trust most for data storytelling feedback between instructor or AI they cited preferring instructor feedback over AI for more assignment-specific feedback. One participant put it this way, “\textit{it depends on what kind of feedback im looking for feels like instructor if im reaching out then i trust they have prior knowledge and experience of what good stories look like they might have their own i want to replicate.}” There was also the idea one participant mentioned that “\textit{the instructor can see what you are doing and can leverage their prior knowledge whereas AI only knows what you show it”} suggesting participants view the instructor as possessing more relevant contextual knowledge.
They noted specific features, like the AI-generated feedback as particularly helpful for more technical support like confirming quick facts or providing narrative refinements. One participant also noted that “\textit{AI is great for asking multiple questions compared to instructor where they may be hesitant to ask everything out of fear of being annoying and also without feeling like need to be as coherent.}” Some participants also highlighted wanting clear distinctions in the interface between AI and instructor feedback. 
While they expressed the perceived usefulness of the tools affordances, several participants proposed more clarity around how to interact with the tool. Since some of the students had already used the tool for an assignment, they were not seeing this version Story Studio for the first time. Still, the tutorial page was a surprise to many, with some users saying they did not see it the first time they used it. They highlighted the need for a more concise tutorial, and some noted the utility of having affordances clearly labeled within the context (i.e. a description below each section stating what actions can be taken). These findings indicate a need for more scaffolding or clarity around the data story construction affordances of the tool. 
%\subsubsection{Narrative Feedback Experience}Participants reviewed preliminary features around narrative templates and envisioned that the feedback would be tailored to to their dataset as well as their audience. One participant put it this way “\textit{these are good narratives, but you need to find a perfect place to apply those}” highlighting that not only do they need help knowing how to structure their data story but also where it could be targeted. They also mentioned the need to balance scaffolding and giving students agency to learn how to modify the templates themselves. One participant noted that,"\textit{if it teaches you to build something a certain way and if it’s not applicable every time then the student also can get confused because it’s just how they learned how to approach it}", suggesting that scaffolding should be done in a way where students can still adapt what they learned to other contexts. Another said it was important to let students know why particular examples are used. 
%\subsection{Analysis}
These insights, coupled with our analysis of feedback frameworks in literature inform the design of Study II feedback alternatives (requirement, timing, locus). We detail these in the next section.

\section{Study 2: Design of Feedback Strategies for Story Studio}
\label{section: study2}

\subsection{Study Objectives and Research Questions}
The main objective of this study is to evaluate the efficacy of feedback strategies incorporated in Story Studio: an AI-assisted data storytelling platform developed at the University of Alabama. We aim to address the research question:
\textit{What is the efficacy of user-preferred feedback modes for AI-assisted data storytelling tools?} (RQ2)

\subsection{Story Studio Key Features}

\begin{figure*}[t]
  \centering
  \includegraphics[width=0.9\textwidth]{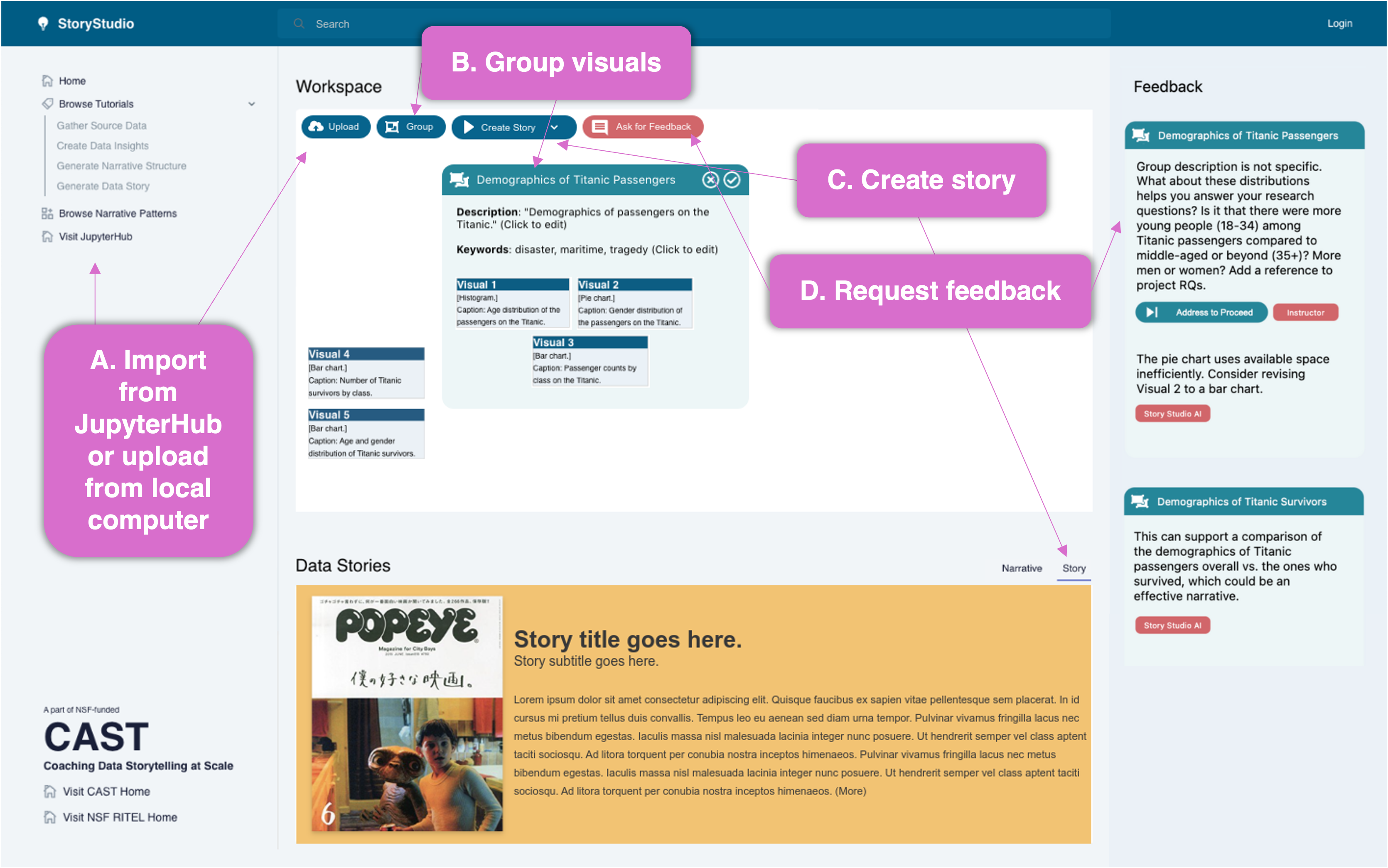}
  \caption{\new{Story Studio website features. (Top-left) Story Studio website main page featuring workspace with data plots. (Bottom-left) Narrative story suggestions. (Top-right) Address feedback feature showing feedback on data story and button "address to proceed".}}
  \Description{This shows the Story studio website and the main workspace. The left hand menu navigation shows different options including Browse tutorials and Narrative Patterns. It also includes a link to return to JupyterHub. The top of the main body displays  buttons horizontally for uploading plots, grouping plots, creating a story, and asking for feedback. Beneath, there is a workspace with a modal for placing plots into a group. Next, the bottom panel of  the main body shows narrative story suggestions. The right hand column displays feedback on the data story, each with a button to address the respective feedback to proceed. Labels are included with the feedback as well, indicating whether it is instructor or Story Studio based feedback. }
  \label{fig:ss_features}
\end{figure*}

\textbf{Story Studio} (figure \ref{fig:ss_features}) is an AI-assisted data storytelling platform developed at the University of Alabama. \old{It features an online application connected to JupyterHub server which allows users to pull visualizations from Jupyter Notebooks and import to Story Studio's \textbf{Workspace}. Users can arrange these visualizations into meaningful groups, \textbf{generate captions} and \textbf{select narrative structures using AI}, and \textbf{craft a data story}}.

\new{It provides the following functionalities:}
\new{
\begin{itemize}
\item \textbf{A. Import or upload visuals}: Connect a Jupyter notebook to import visuals, or upload visuals from a local machine
\item \textbf{B. Group visuals}: Collect visuals with a common theme or purpose into a group, and add a title and description
\item \textbf{C. Create story}: Select, manually or using AI, from a list of narrative structures (cause-and-effect, question-and-answer, overview to detail, factor analysis, timeline). StoryStudio generates a story using all Workspace visuals adhering to the selected narrative structure. The final story appears under ``Data Stories" in the ``Story"" tab, with narrative justification in the ``Narrative'' tab. Users can also view a selection of sample data-driven stories in the narrative selection step. 
\item \textbf{D. Request feedback}: Request feedback on the quality of user annotations and groups
\end{itemize}
}

% \begin{figure}[t]
%   \centering
%   \includegraphics[width=\linewidth]{cast_features_red.png}
%   \caption{\old{Story Studio Website Features. (Top-left) Story Studio website main page featuring workspace with data plots. (Bottom-left) Narrative story suggestions. (Top-right) Address feedback feature showing feedback on data story and button "address to proceed".}}
%   \Description{A visual showing study overview.}
%   \label{fig:ss_features}
% \end{figure}

For the feedback system, we evaluated user perceptions of the following design choices: on-demand, automatic, skip, address, process, and outcome. Automatic and on-demand feedback help understand how users solicit feedback and the desired seamfulness during that process (\textit{timing}) \citep{horvitz1999disruption, chalmers2004seamful}. In addition, skip and address feedback target users' reactions to feedback and how they approach accountability to feedback (\textit{requirement}) \citep{shute2008focus, kluger1996effect}. Finally, in examining the process and outcome feedback features, we evaluate users' ability to reflect in action, assess progress relative to goals, and route their attention (\textit{locus}) \cite{hattie2007feedback, kluger1996effect}. 

\subsection{Data Collection and Analysis}
We sent out a survey to gather feedback from users on the features we developed based on insights from Design Workshop 2. Participants rated the systems on a 5-point scale from strongly disagree to strongly agree. They rated the design of 6 different feedback strategies on-demand, automatic, skip, address, process, and outcomes across three dimensions: effectiveness, persuasiveness, and usefulness. Please refer to Appendix section \ref{survey_questions} for a list of survey questions.
A total of n=7 participants respond to the survey including both students and instructors who help us to understand perceptions from current users (students) and also potential users (instructors). We conducted frequency analysis of survey results and thematic analysis of open-response questions. 

%(? We may not need this here) Example question: 
%Do you think this “on-demand” feedback mode is effective, persuasive, useful? (Likert-style) What would you change or amend about this feature?

%(? Is this a dictionary of codes?) This included consideration of concepts like solicitation, seamfulness, reactions to feedback, accountability, nature, and frequency. 

\subsection{Study Findings}

\begin{figure*}[t]
  \centering
  \includegraphics[width=0.9\linewidth]{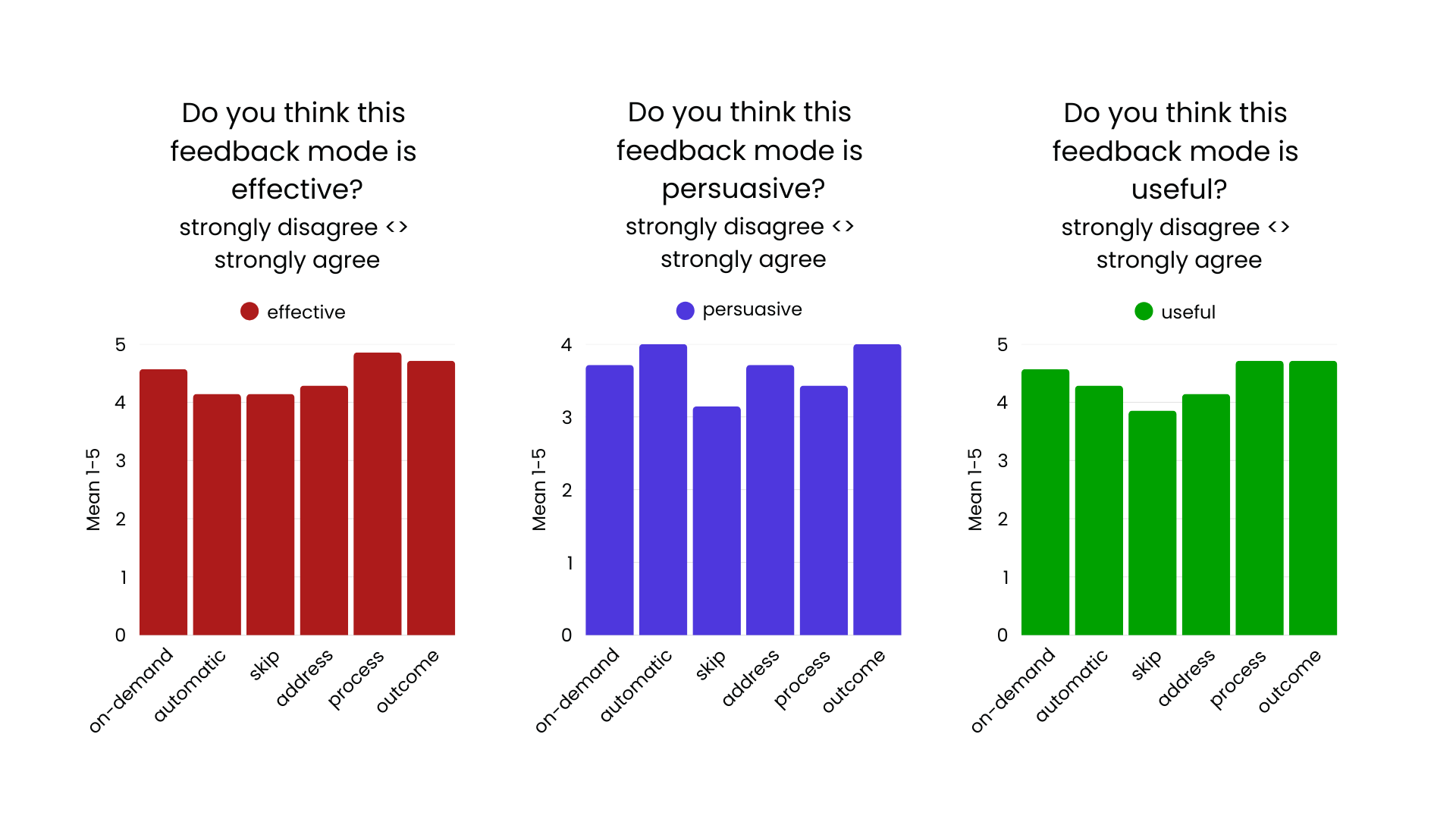}
  \caption{Survey results: (from left) all six feedback alternatives (on-demand, automatic, skip, address, process, outcome) rated on effectiveness, persuasiveness, and usefulness.}
  \Description{A graphic showing three bar charts with results of feedback survey including Likert scale response means around the effectiveness, persuasiveness, and usefulness of the feedback modes: on-demand, automatic, skip, address, process, outcome) each bar chart showing 6 bars. Axis labels include x-axis as the feedback modes and y-axis labeled "mean 1-5" between 1 and 5. All feedback modes were rated positively around effectiveness. Persuasiveness had a wider range of responses along with usefulness. Exact values in Survey Analysis Table \label{survey_analysisresults} in appendix. }
  \label{survey_three_aspects}
\end{figure*}

Figure \ref{survey_three_aspects} presents an overview of participants’ perceptions of the feedback modes in terms of their effectiveness, persuasiveness, and usefulness.

\subsubsection{Timing: Automatic Vs. On-Demand Feedback}
\begin{figure}[t]
  \centering
  \includegraphics[width=\linewidth]{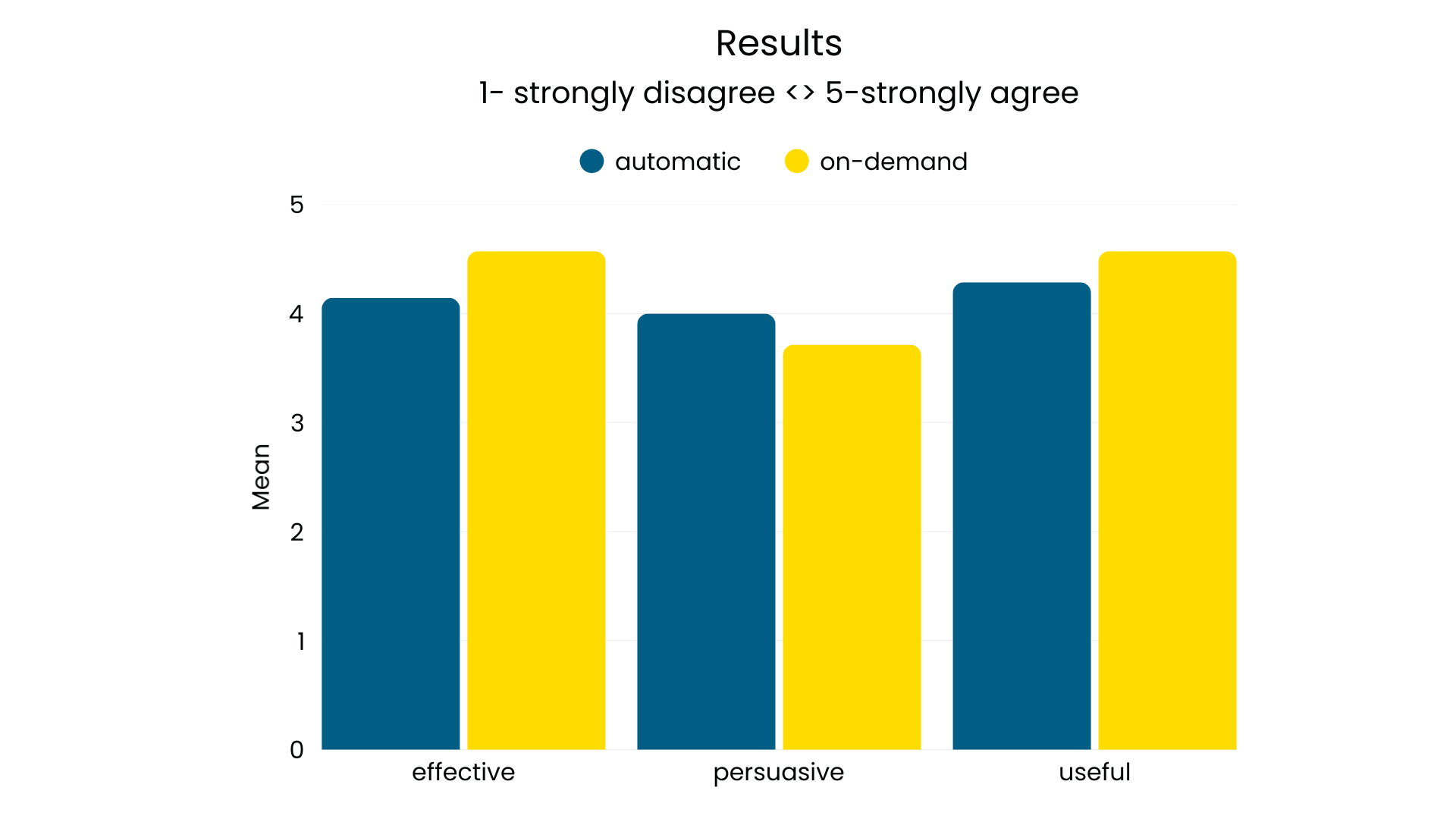}
  \caption{Survey results: automatic vs. on-demand feedback according to effectiveness, persuasiveness, and usefulness.}
  \Description{A graphic showing 3 bar charts  with results of feedback survey for automatic and on-demand feedback dimensions. It shows a blue bar for automatic and a yellow bar for on-demand and the average response across each measure, effectiveness, persuasiveness, and usefulness. Automatic feedback was more persuasive with an average of 4.14 and on-demand was rated as more effective and useful with averages of 4.57 and 3.71, respectively.}
  \label{automatic_on_demand}
\end{figure}

As shown in Fig. \ref{automatic_on_demand}, participants rated on-demand feedback as moderately effective (M=4.57, SD =0.53) as compared to automatic feedback (M=4.14 , SD = 1.06) suggesting automatic feedback might work really well for certain users as opposed to others. Some participants also rated the on-demand feedback positively in the open response saying it was a great feature and wouldn’t change anything, and one even suggested testing the feature as is. 

However, even the on-demand feedback could use some improvement according to open responses given where one participant asked \textit{“could there be a more targeted ask? Like} \textit{'I’m worried about Viz 1, does it work with the others?}'\textit{ Giving students practice asking for specific feedback rather than ‘is this good’ would be helpful.}”  This suggests that soliciting on-demand feedback could be made more useful if it were to target those specific asks based on the context to improve seamlessness. Others mentioned that “\textit{You might want to add a level of importance to the feedback notes (suggested, recommended, essential, …}” suggesting that on-demand feedback also needs to be contextualized according to its importance to determine its effectiveness and usefulness, so that users can determine how much impact the feedback solicited will have on their story. 

Interestingly, although on-demand was rated moderately more effective and useful, automatic feedback was still rated as moderately more persuasive. This suggests that while they are persuaded to use automatic feedback, they might not find it effective or useful in the current design. Some suggestions included providing students the option to turn off the automatic feedback “\textit{so as not to stifle creativity}”, suggesting the lower rating of effectiveness might be tied to perceived barriers to telling a story creatively. This suggests some participants view creativity as an important part of an effective data story. Another participants responded similarly saying that “\textit{I worry that this will result in every story being the same. AI-generated feedback has the potential to provide too strong of a nudge.}” This also speaks to perceived need for balancing solicitation with seamlessness to allow for creativity and originality on the users part. Addressing this may not be a one size fits all as one participant suggested that they can see some students wanting feedback by asking for it and others using automatic feedback instead. They agreed, however, that testing this further would be helpful.

Overall, these results indicate that users feel there needs to be a balance between soliciting feedback and seamless feedback for data storytelling so that stories retain the originality that comes from user input. They suggested that solicitation methods should provide information about how important it is to incorporate specific feedback. The results also indicate that seamlessness should be more around making the system able to answer specific questions on-demand about visuals or components of the story they are referring to, in addition to automatically knowing correct timing for feedback. 

\subsubsection{Requirement: Address Vs. Skip Feedback}
\begin{figure}[t]
  \centering
  \includegraphics[width=\linewidth]{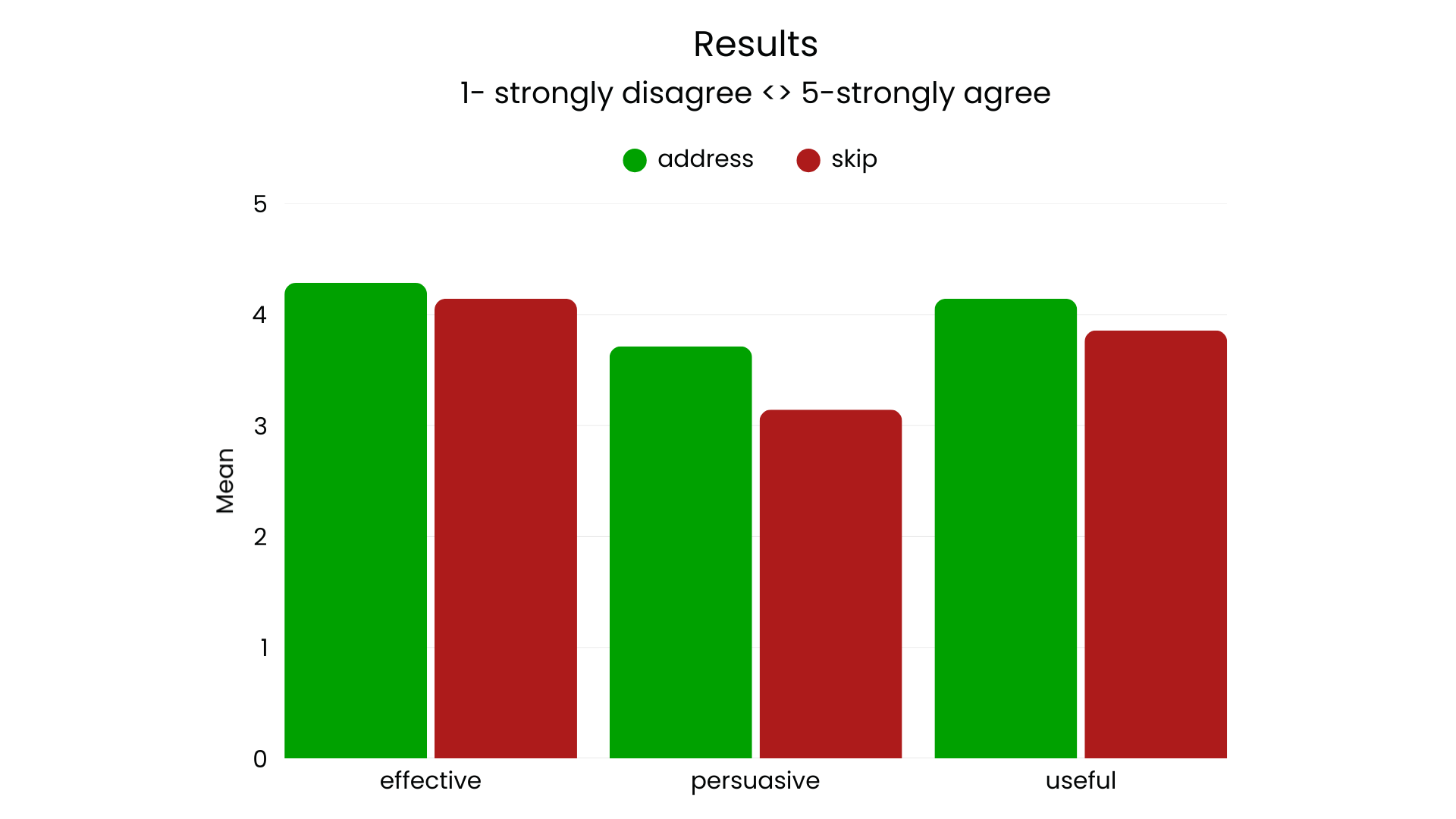}
  \caption{Survey results: address vs. skip feedback according to effectiveness, persuasiveness, and usefulness.}
  \Description{A graphic showing 3 bar chart showing results of feedback survey for address and skip feedback dimensions. It shows a green bar for address and a red bar for skip and average response across each measure, effectiveness, persuasiveness, and usefulness. It shows address as more effective, useful and persuasive compared to skipping feedback.}
  \label{address_vs_skip}
\end{figure}

As shown in Fig. \ref{address_vs_skip}, participants found addressing feedback as more effective, persuasive and useful than the skip feedback feature however there was less agreement over its persuasiveness and usefulness. It was also rated moderately persuasive as compared to other features.  
One participant put it this way, “\textit{I like that this forces revision before proceeding and slows down the "let's just get this done" mentality.}” and another echoed this sentiment stating “\textit{the way that this features ensures that the student doesn’t just skip over the feedback is very smart and especially persuasive}”. Another participant said they were “\textit{unsure on how effective this would be. I find that with suggestions can be quite narrow from AI assistance to a point where it doesn't allow for wiggle room with things, so to have AI assistance have a point to 'require' changes may not be the best}." While some respondents mentioned that they thought it was fine to skip feedback and they like giving students the option, one instructor felt that it was still important to “\textit{insist on students looking at the feedback in a meaningful way}”. This speaks towards incorporating some type of accountability into feedback. To achieve this one participant suggested a limited number of skips. Interestingly, one student mentioned how they would like to “\textit{move skip feedback button earlier so students aren’t influenced by unwanted feedback}”, suggesting that there is a certain timing for when they want to react to feedback so that it does not overly influence the resulting product. 

Taken together, these results suggest that respondents felt requiring accountability is useful and important for learning but it is also important to consider whether enforcing it too often will put constraints on the creative process. 

\subsubsection{Locus: Outcome Vs. Process Feedback}
\begin{figure}[t]
  \centering
  \includegraphics[width=1.0\linewidth]{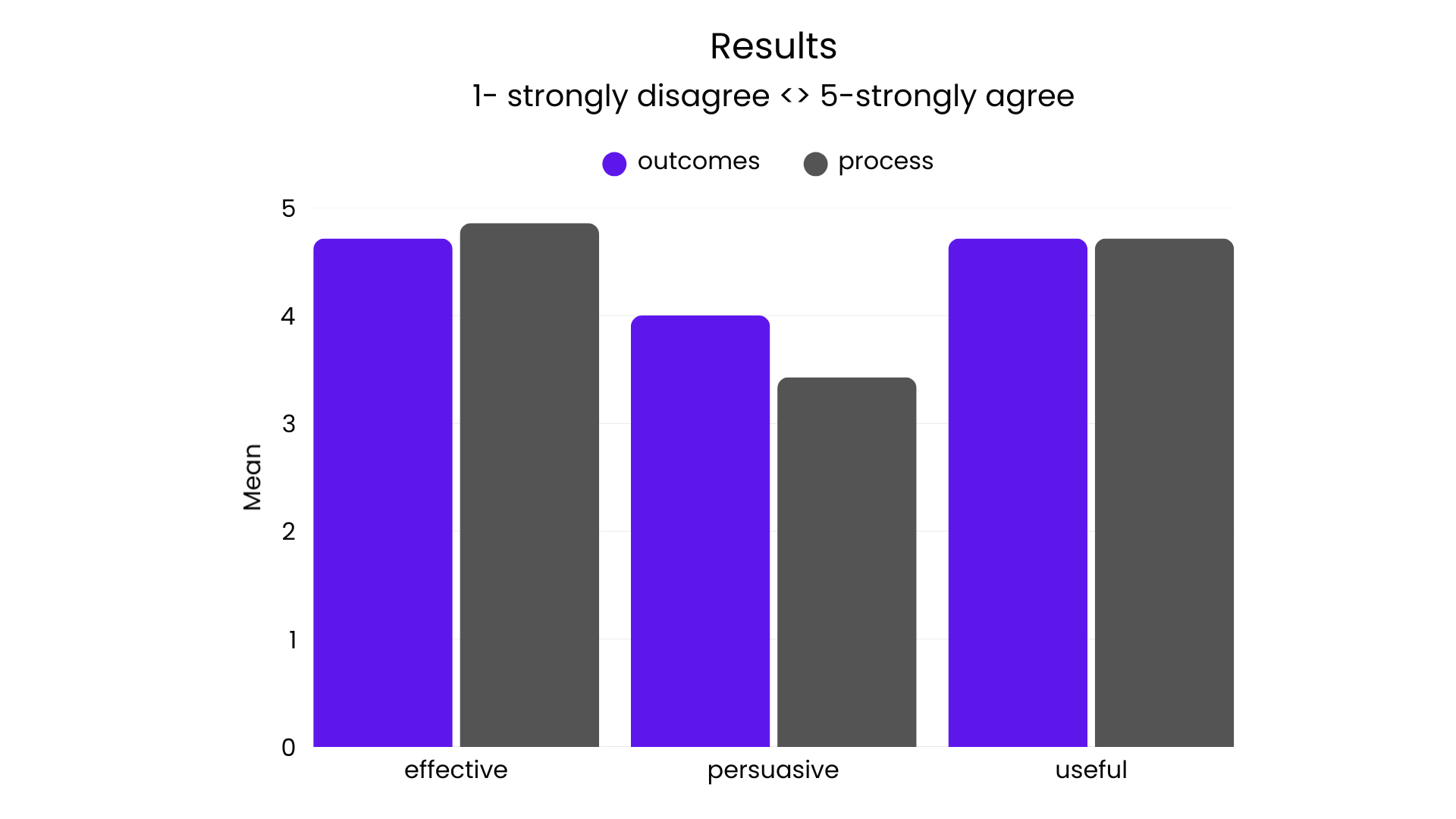}
  \caption{Survey results: outcomes vs. process feedback according to effectiveness, persuasiveness, and usefulness.}
  \Description{A chart showing results of feedback survey for feedback dimensions outcome and process.It shows a purple bar for outcomes and a gray bar for process and average response across each measure, effectiveness, persuasiveness, and usefulness. Process feedback was rated more effective than outcomes feedback. Outcomes feedback was rated as more persuasive than process feedback but equally useful.}
  \label{outcome_vs_process}
\end{figure}

As shown in Fig. \ref{outcome_vs_process}, participants agreed on rating process feedback as moderately effective but felt that the outcomes feature was moderately persuasive (M = 4 , SD = 1.154700538). As compared to the process feature, there seemed to be a more neutral outlook even if participants rated it as more effective. 
In terms of usefulness, they rated both process and outcome feedback as equally useful even if they differed on whether features were effective or convincing enough to actually use. Some respondents perceived process feedback features like “whats next” as especially helpful but with suggestion “\textit{keeping the feedback about ‘what's next’ general rather than too targeted is important to not overly influence the stories students come up with.}” Overall, they felt it was a good feature. For outcomes feedback there was more disagreement around how effective it would be and one suggested “\textit{some information about the categories that they are being evaluated on would be helpful. I can imagine students asking "why did I only get 5/10?}" on communication for example.” Another participant mentioned they would “ \textit{provide brief explanations for where there is incomplete work. I'm unsure of the numeric grading because I think there's a balance between AI assistance and then what the professor has in mind, which may not link up to the AI assistance to warrant 10/10s unless professor's insights/goals match up 100 percent with the AI assistance.}” This echoes another participants response discussing the need for ensuring a human validates the feedback given. Together these results provide important insights into the effectiveness of these features and what would make them more persuasive and useful to use for learning. In particular, what emerges is another example of ensuring the feedback does not overly influence the narrative and also ensuring that it aligns with the instructors expectation for how outcomes should be measured. 

\section{Discussion}
\new{In our findings we presented insights from observations, interviews and workshops extending previous work on data storytelling processes \cite{lee2015story,nolan2021communicating,yarnall2025laying}. Through an analysis of the cognitive tasks and challenges that occurred during the process and the ideated solutions for potential feedback features, we provide support for our first research question (RQ1): What are learners and educators’ needs and expectations of feedback on data storytelling workflows? We extend \cite{yarnall2025laying} work by using it as a lens for analysis of these needs and expectations and use it as a framework for the design space of data storytelling feedback. Their work highlighted the skills and tasks required in the data storytelling process as well as opportunities for designing intelligent feedback to support those skills. Our work is in line with their design recommendations and we extend their use of evidence centered design for this context \cite{yarnall2025laying,Haertel2012ECD} .} 

\new{We also presented findings from our surveys to answer our second research questions (RQ2): What is the efficacy of user-preferred feedback modes for AI-assisted data storytelling tools? These results augment past work on feedback in education and learning sciences as well as work on feedback in creative storytelling \cite{schon1983reflective, csikszentmihalyi1996creativity}. Our findings also augment literature indicating feedback as a strong intervention for supporting learning \cite{hattie2007feedback,Wang2018PublicPeerReview} by aligning our design space with supporting better outcomes.} 
\subsection{Feedback in Data Storytelling: Balancing Accountability with Creativity}
Participants’ mixed response to automatic versus on-demand feedback underscores the challenges of delivering feedback at the right moment in the workflow. On-demand feedback is rated as moderately more effective and useful, with students valuing the control it gives them to request input when ready. This is in alignment with Horvitz \cite{horvitz1999disruption} which emphasizes respecting users’ availability and mental state. Automatic feedback, however, is seen as moderately more persuasive, suggesting that unsolicited nudges can be powerful in prompting engagement. \new{While the results from the workshops showed it could be helpful for real-time feedback on clarity and relevance to audience,p}articipants worried that automatic feedback might stifle creativity or homogenize stories, in line with Chalmers and Galani \cite{chalmers2004seamful} that seams must be staged carefully rather than hidden outright. In creative workflows, seamlessness requires not only unobtrusiveness but also a sensitivity to preserving originality. \new{ However, as indicated in Section 4.3.1, when students are still unsure how to proceed after automatic feedback, seamlessness needs to be supplemented with opportunities for clarification on-demand. }

Findings on the requirement to address feedback highlight the dual role of accountability in encouraging reflection and limiting creative ideation. Many participants \new {in both the workshops and surveys }valued enforced revision, noting that it slowed down the impulse to “just get it done”, and encouraged reflection on the emergent insights and narratives. This resonates with Shute’s framing of formative feedback as most effective when it promotes meaningful engagement \cite{shute2008focus}. Other participants noted that required revisions could be overly narrow in scope, creating over-reliance and limiting creative flexibility. Required feedback can also be seen as a seam: for some learners, it drew attention to reflection, while for others it introduced unwanted friction. This duality is reflected in Chalmers and Kluger's models where feedback can shift attention both productively and disruptively. \new{In this context, required feedback needs to strike a balance as a productive disruption. Required feedback should also provide agency to override suggestions if the student determines the issue has been adequately addressed while also ensuring there is accountability for feedback provided. }

A similar need for calibration emerges in participants' preference for process and outcome feedback. Participants rated process feedback as moderately effective and useful, especially when framed as general “what’s next” guidance rather than highly targeted prescriptions. Hattie and Timperley \cite{hattie2007feedback} observe that process-level feedback drives learning, while overly directive process cues may undermine originality in creative tasks, consistent with our findings \new{that indicated users were concerned about creativity and less inclined to want to engage with extra details especially in narrative feedback.} Outcome feedback, by contrast, was perceived as more persuasive but slightly less effective. \new{Findings from our workshops also indicated that outcome feedback was perceived as ideal particularly for those that did not want overly scheduled feedback or too much detail in the scaffolding for narrative structures provided and preferred a one-time grade to reflect over at the end.} The distinction between reflection in-action and on-action observed by Schön \cite{schon1983reflective} provides an explanation for this difference: process feedback supports in-action reflection while outcome feedback struggles to support meaningful on-action reflection without contextualization or human validation. According to FIT \cite{kluger1996effect}, outcome feedback redirects attention toward grades and evaluation as opposed to the storytelling tasks. Our findings identify the need for flexible in-action feedback, lightweight nudges, and transparency about grading rubrics and instructional goals.
\new{Our results identify perceived effectiveness of AI feedback for this process that extends work done in the field of learning sciences exploring the role of feedback in learning, particularly perceptions around feedback and their effect on the propensity to revise and success at increasing the quality of writing \cite{Wang2018PublicPeerReview,Cheah2020StructuredFeedback}. The findings also provide support for other work that highlights structured feedback as effective for learning in the context of educational design \cite{BangertDrowns1991Feedback, Ilgen1979Feedback,Timmers2015MotivationFeedback} and for improving writing quality \cite{graham2011informing,macarthur2016instruction,panadero2023effects}}

\subsection{Implications for Design of AI-Augmented Data Storytelling Systems}
\new{Our work builds on work by \cite{kosara2013storytelling} that focuses on data storytelling process and tools that support this process \cite{Ren2023DataStorytellingNarrative,shi2020calliope,Zhao2021Chartstory,Wang2018Narvis} and extends it by providing recommendations for how to implement AI  feedback to support learning.} Based on the discussion in preceding sections, we identify the following best practices for design of AI-augmented storytelling systems:
%\begin{figure}[t]
%  \centering
  %\includegraphics[width=0.75\linewidth]{designimplicationtable.png}
 % \caption{\new{Implications for design of AI-mediated data storytelling support systems.}}
%  \Description{A chart showing design implications of feedback dimensions timing, requirement, and %locus. Design implications include Adapting to Storytelling stages, Avoiding Over-scheduling, and Clarifying Roles and Credibility.  }
%  \label{design_implications}
%\end{figure}

\begin{table*}[t]
\centering
\small
\setlength{\tabcolsep}{6pt}
\begin{tabular}{|p{4cm} |p{4.2cm} |p{4.2cm} |p{4.2cm}|}
\hline
\textbf {Design Implications} &
\textbf{Timing: On-demand vs Automatic} &
\textbf{Requirement: Address vs Skip} &
\textbf{Locus: Process vs Outcome} \\
\hline
Adapt to Storytelling Stages &
automatic early stages, on-demand depending on feedback target &
require in early stages, allow optional skip later &
support general reflection-in-action across stages \\\hline

Avoiding Overscheduling &
scheduled at key transition points &
user awareness and accountability with constraints &
control feedback frequency during process \\\hline

Clarifying Roles and Credibility &
automatic feedback should provide clear justification &
require addressing feedback aligned with instructor &
outcome feedback should be instructor backed \\
\hline
\end{tabular}
\caption{Design implications across feedback timing (on-demand vs automatic), requirements (address vs skip), and locus (process vs outcome).}
\Description{Table showing three design implications and how they map to timing of feedback, whether feedback must be addressed or skipped, and whether feedback focuses on process or outcome.}
\label{tab:design-implications}
\end{table*}

\begin{itemize}
    \item \textbf{Adaptive requirement}: \old{AI data storytelling systems} \new{AI-enabled data storytelling assistants} should balance accountability with creative agency, by varying requirement across different stages of the workflow. Early stages of the workflow (crafting research questions, grouping visuals and claims into insights) can benefit from requiring engagement with feedback to inspire reflection. Later stages may benefit from optional or limited-skip controls that preserve creative flexibility. \new{Process feedback should remain general enough to support reflection-in-action without over-constraining narrative direction.These adaptive requirements can address the need for flexible interaction in creative multi-stage workflows that shift from exploratory and convergent phases \cite{csikszentmihalyi1996creativity} by structuring feedback systematically according to task stage or type of reflection. }

    \item \textbf{Avoiding over-scheduling}: AI-assisted feedback should be scheduled at key transition points in the data storytelling workflow, with affordances to maintain users' state awareness and control of frequency. \new{This approach can mitigate how feedback timing can influence engagement with feedback \cite{horvitz1999disruption} }

    \item \textbf{Clarifying Roles and Credibility:} \old{Process feedback should remain general enough to support reflection-in-action without over-constraining narrative direction, while outcome feedback should be contextualized with clear justifications and course staff's credibility cues to build trust.} \new{AI- assisted outcome feedback should be contextualized with clear justifications and course staff's credibility cues to build trust. The role of the AI assistant should be clarified so that the user understands where they will provide feedback and where the instructor provides the feedback}. 
\end{itemize}

\subsection{Limitations and Future Work}
We review key limitations of our study and associated directions for future work as follows. Our study participants are faculty and students affiliated with the Information Systems department at a single institution, which limits generalizability of our \new{quantitative} findings. \new{While sample size considerations were made within the range of qualitative studies, they were on the lower end \cite{Caine2016LocalStandards,Bacchetti2011BreakingFree,PloutzSnyder2014SmallN,Tisdell2025QualitativeResearch} and it is possible we did not capture full range of perspectives among data storytelling students.} We plan to conduct multi-institution studies to better assess the effectiveness of AI-mediated feedback in data storytelling. \new{We also acknowledge that feedback preferences may be separate from what actually supports learning.} The scope of our evaluation is exploratory. In future work, we plan to evaluate Story Studio's feedback efficacy in its impact on time-on-task and user ability for critical reflection, and control for participants' proficiency with data science and narrative storytelling. \new{We plan to incorporate A/B testing to assess learning outcomes more rigorously. More work should be done to explore ways to assess data storytelling skill acquisition given it is an ill-defined domain which can be complex to operationalize. } \new{While this study does not explicitly look at student learning as an outcome measure, these findings nonetheless contribute to a larger vision that supports learning by providing an understanding of student's judgment of AI feedback. By contributing this understanding of user's perception of the technology we provide a necessary starting point for future learning outcome research by focusing on what student's perceive as useful first to avoid measuring learning outcomes on a system design that does not align with those perceptions. Finally, we also hope to expand our investigation to data storytelling with immersive media, decision-assistance platforms, and resource-constrained networked information systems \cite{sajid2025just, hassan2024simplify, khalid2014home}.}

\section{Conclusion}
Our study explores the design of AI-mediated feedback for data storytelling workflows. Through surveys, interviews, and design workshops, we surface user preferences for feedback scheduling, requirement, and attentional locus. We find that adaptive requirement, user agency, and credibility cues can improve the balance of accountability and creative flexibility in these complex, multi-step workflows. These preliminary findings provide support for existing work on the importance of perception of feedback in relation to its effectiveness in the context data storytelling education. They also define a potential direction for this design space and future work when seeking to understand students' and instructors' expectations for AI-mediated feedback for data storytelling.  

\section{Acknowledgment}
This research was supported by the National Science Foundation under Grants Nos. 2302794 and 2302795. Any opinions, findings, conclusions, or recommendations expressed in this material are those of the authors and do not necessarily reflect the views of the foundation.

\appendix

\section{Study Instruments}
\subsection{Experts}
\begin{enumerate}
\item What types of timely feedback do you think would be most important to provide to entry-level data scientists to help them adjust and improve their processes of data questioning, data exploratiion and analysis, and data storytelling? 
\item What types of dashboard features (e.g., student-facing coaching hints, or instructor-facing feature to provide feedback to students) would enhance your training strategies in this tool? 
\item What aspects of asssessing data storytelling processes and products should be automated using this tool? Which should not? Please share some rationale to help me understand your preferences.
\item What are easiest and hardest for data scientist to develop data stories? Why?

\end{enumerate}
\subsection{Students}

\begin{enumerate}

\item What is your story about?
\item  What were your initial questions after knowing about the data?
\item  Did your data story answer all of your initial questions? 
\item What steps did you take to develop your story?
\item What was the easier part while you developed the story?
\item What was the hardest part while you developed the sotry?
\item What kinds of feedback did you receive or have liked to receive from your instructors? 
\item Did you have time to act on that feedback and revise your data story? If so, please briefly describe that process. If not, please briefly describe what you would have liked to adjust to improve the story. 
\item What features would be helpful in an AI tool? 
\end{enumerate}
\subsection{Design Workshops}
\subsubsection{Design Workshop I}
\label{design_workshop1}
Brainstorming Prompts Based on Streaming Scenario:
\begin{enumerate}
    \item "What kinds of challenges do you think Jamie might face when analyzing data for such a varied audience?"
    \item "How might the assistant help in customizing insights based on different audiences, and what specific features would make this process smooth?"
    \item "Would it be valuable if the assistant could detect and highlight insights that are most relevant for decision-making? How could this feature work?"
    \item "What specific types of feedback or guidance would be useful for Jamie in refining key takeaways for a diverse audience?"
\end{enumerate}

Brainstorming Prompts Based on Healthcare Scenario:
\begin{enumerate}
    \item "What kinds of challenges do you think Alex might face when analyzing data for such a varied audience?"
    \item "How might the assistant help in customizing insights based on different audiences, and what specific features would make this process smooth?"
    \item "Would it be valuable if the assistant could detect and highlight insights that are most relevant for decision-making? How could this feature work?"
    \item "What specific types of feedback or guidance would be useful for Alex  in refining key takeaways for a diverse audience?"
\end{enumerate}

Groups 2 and 3 were shown results from interviews and how they said they would want support. 

\begin{itemize}
    \item Ask what stage in process you are of story
    \item Provide recommendations for what to do in each stage 
    \item “Look” at visualization and ask specific questions 
    \item General guidance if user doesn't know 
    \item Answer specific questions if user already knows 
    \item Step by step for data analysis for when trying to develop question 
    \item Find interesting questions to differentiate yourself
    \item Understanding context
\end{itemize}

\subsubsection{Design Workshop II}
\label{design_workshop2}
(Format and Appearance)
\begin{enumerate}
    \item Would you rather see feedback as comments, checklists, scores, highlights, or something else?
    \item Do you prefer feedback on the side or as annotations?
    \item How helpful would visual cues like red/yellow/green icons be?
\end{enumerate}

(Timing)
\begin{enumerate}
    \item When is the best time to get feedback? after submission, in stages or at the end?
    \item If you could get it during the process is it helpful or distracting?
\end{enumerate}

(Feedback Actionability)
\begin{enumerate}
    \item Would you prefer suggested next steps or leave it open-ended?
    \item How helpful would it be to have buttons next to feedback items like ‘revise now’ , ‘ask question’ or ‘mark as done’ ?
    \item Would you use a feature that lets you track how you revised something?
\end{enumerate}

(Feedback source)
\begin{enumerate}
    \item Which feedback do you trust most, instructor or AI assistant?
    \item How would you want feedback from the instructor?
    \item Is it different from feedback from an AI assistant?
\end{enumerate}

\subsection{Survey}
\label{survey_questions}
Do you think this (feedback mode) is (effective, persuasive, useful): 
\begin{itemize}
    \item Feedback mode: on-demand, automatic, skip, address, process, outcomes
    \item (strongly disagree <> strongly agree, 5-point Likert scale)
    \item What would you change or amend about this feature? [Response text]:
\end{itemize}
\new{\subsection{Gantt Chart}}
\new{This shows a Gantt Chart visualization we created for one of the groups. It shows the event start and end time, interval, and the different design patterns we labeled the students to be in (DP1, DP2, DP3). It also has a column showing the problem solving steps we observed over time.} 
\label{Gantt Chart}
\begin{figure}[H]
  \centering
  \includegraphics[width=\linewidth]{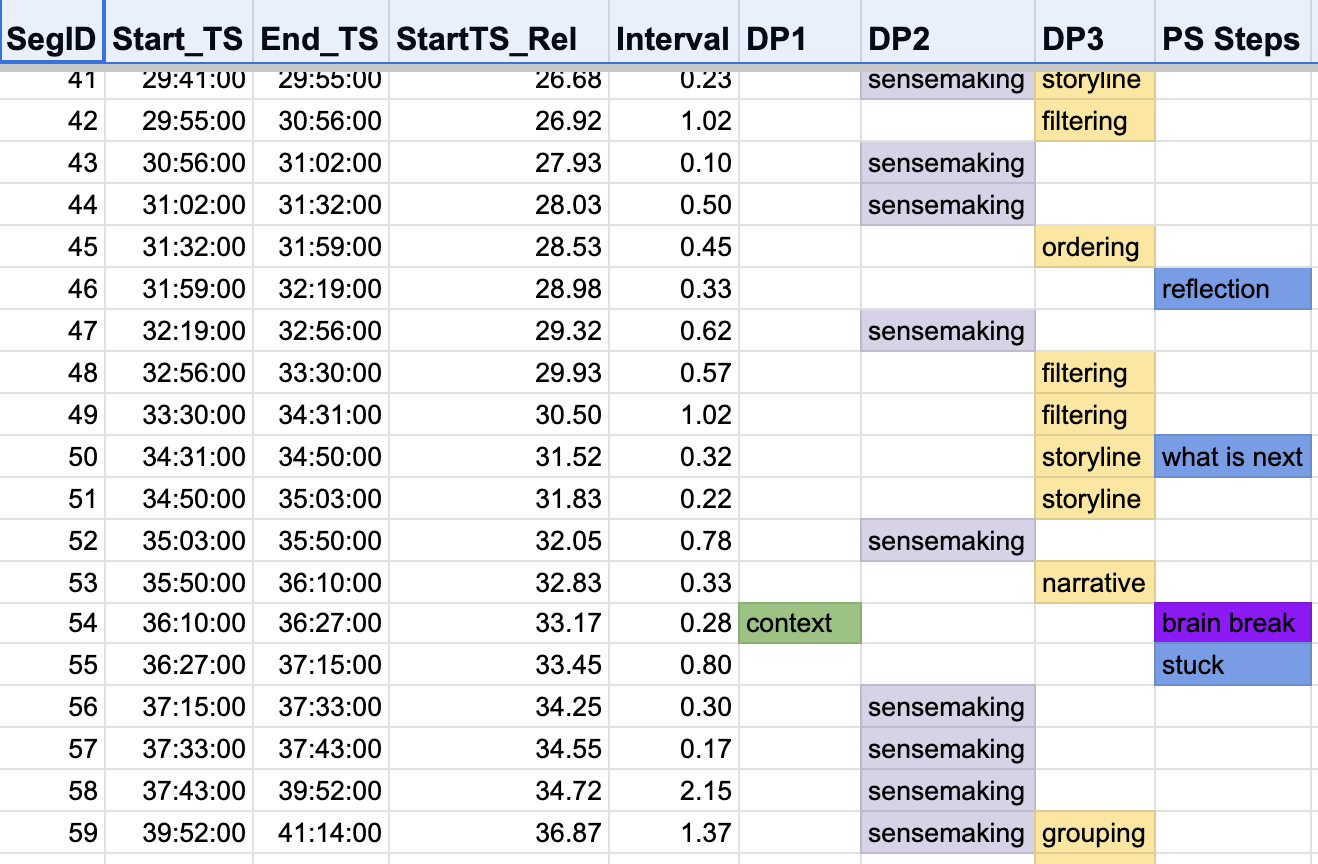}
  \caption{ Gantt Chart overview.}
  \label{Overview Gantt Chart}
\end{figure}
\subsection{Survey Analysis}
\label{survey_analysisresults}
This shows Survey Analysis results of mean and standard deviation. 
\begin{table}[H]
\centering

\begin{tabular}{l l l l}\toprule
\textbf{feature} & \textbf{dimension} & \textbf{average} & \textbf{std dv} \\\midrule
on demand feedback & effective & 4.57& 0.53\\
automatic feedback & effective & 4.14& 1.07\\
skip & effective & 4.14& 0.90\\
address & effective & 4.29& 1.25\\
process & effective & 4.86& 0.38\\
outcomes & effective & 4.71& 0.49\\
on demand feedback & persuasive & 3.71& 1.11\\
automatic feedback & persuasive & 4.00& 0.82\\
skip & persuasive & 3.14& 0.90\\
address & persuasive & 3.71& 1.25\\
process & persuasive & 3.43& 1.13\\
outcomes & persuasive & 4.00& 1.15\\
on demand feedback & useful & 4.57& 0.53\\
automatic feedback & useful & 4.29& 1.50\\
skip & useful & 3.86& 0.90\\
address & useful & 4.14& 1.21\\
process & useful & 4.71& 0.49\\
outcomes & useful & 4.71& 0.49\\ \bottomrule

\end{tabular}
\caption{Table presenting mean and standard deviation of participant responses for the six feedback modes on-demand, automatic, skip, address, process, and outcomes. across measures effectiveness, persuasiveness, and usefulness.}
\label{tab:my_table}
\end{table}

\bibliographystyle{ACM-Reference-Format}
\bibliography{story,feedback}

\end{document}